\documentclass[journal]{IEEEtran}

\usepackage{calc}
\usepackage{stfloats}
\usepackage{calc}
\usepackage{microtype}
\usepackage[intlimits,sumlimits,namelimits]{amsmath} 
\everymath{\displaystyle} 
\usepackage{amssymb}
\usepackage{amsfonts}
\usepackage{psfrag}
\usepackage{verbatim}
\usepackage{multirow}
\usepackage{multicol}
\usepackage{algorithm2e}
\usepackage{blkarray}
\usepackage{wrapfig}
\usepackage{paralist}
\usepackage{trfsigns}
\usepackage{version}
\usepackage{wasysym}
\usepackage{bm}
\usepackage{color}
\usepackage{xcolor}
\usepackage{graphicx}
\usepackage[most]{tcolorbox}
\usepackage[export]{adjustbox} 
\usepackage{caption}
\usepackage[
    backend=biber,
    style=numeric,
    doi=false,
    isbn=false,
    url=false,
    eprint=false,
    sorting=none
]{biblatex}
\renewbibmacro{in:}{}
\usepackage{catchfile}
\usepackage{pgffor}
\usepackage{mathtools}
\usepackage{bigints}
\usepackage{xfrac}
\usepackage{ifthen}
\usepackage{dsfont}
\usepackage{flushend}

\def\vec#1{\underline{#1}}
\def\mat#1{{\mathbf #1}}

\newcommand{\rvec}[1]{\ensuremath{{\boldsymbol{\underline{#1}}}}}

\def\1_2{{\frac{1}{2}}}

\def\etavec{\vec{\eta}}

\def\d{\mathrm{d}}

\def\NewR{{\rm I\hspace{-.17em}R}}

\def\NewP{\mathrm{I\kern-0.15em P}}

\newcommand{\IN}{\ensuremath{\mathds{N}}}
\newcommand{\IR}{\ensuremath{\mathds{R}}}

\def\Eq#1{(\ref{#1})}

\def\Sec#1{Sec.~\ref{#1}}
\def\SubSec#1{Subsec.~\ref{#1}}

\newlength\EqLen

\def\ScaleInner#1{%
  \settowidth{\EqLen}{#1}
  \ifdim\EqLen < \columnwidth%
    \begin{equation*}%
      \begin{minipage}{\EqLen}#1\end{minipage}%
    \end{equation*}%
  \else%
    \begin{equation*}%
      \resizebox{0.99\columnwidth}{!}{\begin{minipage}{\EqLen}#1\end{minipage}}%
    \end{equation*}%
  \fi%
}%

\def\Scale#1
{
  \ScaleInner{$ #1 $}
}

\def\LongVersion#1{}

\def\citep#1{(\cite{#1})}

\usepackage{amsthm}



\newtheorem{remarkx}{Remark}[section]
\newenvironment{remark}
{\remarkx}
{\endremarkx}

\makeatletter
\newcommand\SaveEquation[2]{\@namedef{equation@#1}{#2}}
\newcommand\UseEquation[1]{\@nameuse{equation@#1}}
\makeatother

\SetKw{KwBy}{by}

\SetCommentSty{mycommfont}

\newtcolorbox{YellowBox}{
  enhanced,
  boxrule=0pt,frame hidden,
  borderline east={1mm}{0pt}{yellow!90!black},
  borderline west={1mm}{0pt}{yellow!90!black},
  colback=yellow!40!white,
  sharp corners,
  grow sidewards by=1.5mm,
  top=0mm,
  bottom=0mm,
  left*=0mm,
  right*=0mm
}

\date{}

\author{Uwe D.\ Hanebeck, \IEEEmembership{Fellow, IEEE}
\thanks{Uwe D.\ Hanebeck is with the Intelligent Sensor-Actuator-Systems Laboratory (ISAS), Institute for Anthropomatics and Robotics, Karlsruhe Institute of Technology (KIT), Germany. (e-mail: Uwe.Hanebeck@kit.edu).}}

\title{Newton-Flow Particle Filters\\based on Generalized Cram{\'e}r Distance

}

\usepackage[acronyms,nopostdot,nonumberlist,nogroupskip]{glossaries}

\makeglossaries

\newcommand*{\Acro}[4][]{%
    \ifthenelse { \equal {#1} {} }%
    { \newacronym{#2}{#3}{#4} }%
    { \newacronym[#1]{#2}{#3}{#4} }%
}

\Acro[\glsshortpluralkey={DMDs},\glslongpluralkey={Dirac mixture densities}]{DMD}{DMD}{Dirac mixture density}
\Acro{GM}{GM}{Gaussian Mixture}
\Acro[\glslongpluralkey={mixture densities}]{MD}{MD}{mixture density}
\Acro[\glslongpluralkey={Gaussian mixture densities}]{GMD}{GMD}{Gaussian mixture density}
\Acro{pdf}{pdf}{probability density function}
\Acro{ODE}{ODE}{ordinary differential equation}
\Acro{NFPF}{NFPF}{Newton-flow particle filter}
\Acro{SDE}{SDE}{stochastic differential equation}

\begin{document}

\maketitle
\thispagestyle{empty}
\pagestyle{empty}


\begin{abstract}
    We propose a recursive particle filter for high-dimensional problems that inherently never degenerates. The state estimate is represented by  deterministic low-discrepancy particle sets. We focus on the measurement update step, where a likelihood function is used for representing the measurement and its uncertainty. This likelihood is progressively introduced into the filtering procedure by homotopy continuation over an artificial time. A generalized Cramér distance between particle sets is derived in closed form that is differentiable and invariant to particle order. A Newton flow then continually minimizes this distance over artificial time and thus smoothly moves particles from prior to posterior density. The new filter is surprisingly simple to implement and very efficient. It just requires a prior particle set and a likelihood function, never estimates densities from samples, and can be used as a plugin replacement for classic approaches.

\medskip


\end{abstract}


\section*{Note}

%
%
The first version of this manuscript was prepared as a basis for discussions during my plenary talk at the
\emph{2025 IEEE International Conference on Multisensor Fusion and Integration (MFI 2025)} at
Texas A\&M University, College Station, Texas on September 3, 2025.

%
%
This slightly modified version was prepared for my plenary talk at the \emph{17th Symposium Sensor Data Fusion} in Bonn, Germany, on November 25, 2025. It includes changes based on comments by Ryne Beeson, Ondřej Straka, Daniel Frisch, and Yiqing Wang. In addition, the distance measure was modified to naturally avoid singularities.

%
%
The next version will include sections on implementation details, state-of-the-art, and an evaluation of the proposed filtering method.

\printglossary[title=Glossary of Terms,type=\acronymtype]

\section*{Notation} \label{Sec_Notation}

Vectors will be indicated by small underlined letters, $\vec{0}$ and $\vec{1}$ are a vector of zeros and a vector of ones, respectively. 
Random values will be denoted by small boldface letters, so random vectors are small boldface underlined letters.
Matrices are denoted by capital boldface letters.
The identity matrix of dimension $N$ is $\mat{I}_N$.

\section{Introduction} \label{Sec_Intro}

\subsection{Estimation Problem to be Solved}

%
%
We consider recursive estimation of the state of a nonlinear dynamic system based on measurements sequentially arriving at discrete time steps.
%
%
The evolution of the system state is described by a process model either in continuous time by a \gls{SDE} or in discrete time by a stochastic difference equation, both usually in state-space form.
%
%
The relation of states to measurements is described by a stochastic measurement model.
%
%
The goal is to continually provide an estimate of the true system state with each new measurement.

%
%
Due to the uncertainties in both process and measurement model, only an uncertain estimate of the state can be provided, which is described by a \gls{pdf}.
%
%
This \gls{pdf} is propagated through the process model (prediction step or time update)
%
%
and updated with a new measurement (filter step or measurement update).

%
%
For performing time and measurement update, specific density representations have to be selected.
%
%
As performing the prediction step for significant nonlinearities is demanding for continuous densities, particle representations are typically selected.
%
%
These simplify the prediction step, but lead to difficulties in performing the filter step.
%
%
The focus of this paper is a new method to execute the filter step.

\subsection{Challenges of Particle-based Estimation}

%
%
The prediction step entails propagating the particles forward through the system model, which is relatively straightforward.
%
%
The filtering step, however, although looking almost trivial at first glance, is far more complex to actually implement.
%
%
It requires finding a particle set compatible with both the prediction and a given measurement. In a Bayesian setting, this results in a multiplication of the prior density and a likelihood function (resulting from measurement equation and actual measurement).
%
%
Naïvely performing the Bayesian update step would result in a set of weighted particles for the posterior density. Especially for ``narrow'' likelihoods, many particles would receive zero weights, and not contribute to the density representation. This effect is called particle degeneration, and many methods for its avoidance have been published.

%
%
Many of the methods for restoring the health of a particle set propose resampling methods that rely on the subsequent prediction step.
%
%
The bootstrap particle filter, for example, just replicates particles according to their relative weight in the filter step without changing their locations.
Only the process noise of the following prediction step is then responsible for introducing particle variability.
%
%
This leads to an undesired dependency of the filter step from the parameters of the process model, e.g., its noise covariance, and is also problematic for chaining several filter steps without having intermediate prediction steps.

%
%
One of our goals for the filter developed in this paper is the strict separation of the filter step from the prediction step: The filter step should be self-contained.

\subsection{New Filtering Method: Key Ideas}

%
%
In this paper, we advocate a radically different method for particle-based filtering that inherently avoids sample degeneration issues.

%
%
It is based on the idea of progressive Bayesian estimation in \cite{SPIE03_HanebeckBriechle-ProgBayes,MFI03_Hanebeck,CDC03_Feiermann-ProgBayes}, which employs continuous densities for representing the posterior.
Progressive estimation for \glspl{DMD} was introduced in \cite{Fusion11_Ruoff,FUSION23_Hanebeck}, where direct discretization of the artificial time related to the homotopy was used, resulting in a discrete flow.
In contrast, in this paper, \glspl{ODE} for continuous particle flows will be derived.
Compared to \cite{arXiv18_Hanebeck} the set distance measure from \cite{AT15_Hanebeck} is used.

%
%
The method consists of three main ingredients that will be explained in detail in the following:
(i)~progressively introducing the likelihood by homotopy continuation,
(ii)~deterministic sampling as density representation,
and (iii)~using a set distance measure for deriving a Newton particle flow from prior to posterior density.

\paragraph{Homotopy Continuation}
%
%
Instead of performing the filter step at once with the given likelihood function, an artificial time $\gamma$ ranging between $0$ and $1$ is introduced, which is then used for parametrizing the likelihood. For $\gamma=0$, no update is performed. For $\gamma=1$, the original likelihood is recovered.
%
%
The posterior density is also parametrized with $\gamma$ so that it starts as the prior density for $\gamma=0$ and moves towards the (unknown) posterior for $\gamma \rightarrow 1$.
%
%
For a particle representation, this so-called homotopy continuation approach results in a particle flow over the artificial time $\gamma$.

\paragraph{Deterministic Sampling}
%
%
For the density representation by particles, we use deterministic samples (also called low-discrepancy samples).
%
%
Compared to random samples,
(i)~the density is covered more homogeneously,
(ii)~the number of particles required for coverage is reduced,
(iii)~statistical properties, e.g., moments, show a much better convergence,
(iv)~and reproducibility is ensured.
%
%
In this paper, we use the sampling method from \cite{hanebeckDiracMixtureApproximation2009} also used in \cite{steinbringSmartSamplingKalman2016}.

\paragraph{Newton Flow Induced by Set Distance Measure}
%
%
The flow induced by the homotopy continuation very rarely produces exact results.
In general, we have to be content with approximations.
%
%
This means that we have to continually track the closest approximation to the true posterior over the artificial time.
%
%
Closeness is measured by a set distance measure, the so called generalized Cram{\'e}r-von Mises distance \cite{AT15_Hanebeck,arXiv14_Hanebeck-Reduction} or Cram{\'e}r distance for short, that allows the efficient comparison of \glspl{DMD} and is differentiable w.r.t.\ to the particle locations and weights.
%
%
The optimal particle approximation is tracked over artificial time $\gamma$ by a Newton flow induced by this generalized Cram{\'e}r distance.

\subsection{New Filtering Method: Unique Features}

%
%
Here, we will give a short overview of the advantages of the new filtering method and also of its differences w.r.t.\ the state-of-the-art.

\paragraph{Modest Assumptions}
%
%
For performing the filter step, the new method only requires (i)~the prior density in the form of a (possibly weighted) particle set and (ii)~a likelihood function.
Nothing else is required!
%
%
In particular, the underlying continuous prior density is not required and no attempt towards its reconstruction is ever made.
%
%
This is the natural setup encountered in recursive particle filtering, where the last estimate is a particle set used in the next filter step.
%
%
Avoiding density reconstruction is especially important in higher-dimensional spaces as density estimation from samples is notoriously difficult here.

\paragraph{Mapping-free Flow}
%
%
The new filtering method is different from other particle flow filters in that it does not calculate an explicit mapping for transporting particles.
%
%
This can be interpreted as a direct online optimization of the distance measure versus finding a mapping policy first.
%
%
Such a policy is often found by interpreting the homotopic version of the measurement update equation as a Fokker-Planck equation and using the associated \gls{SDE} for moving the particles.
%
%
This necessitates first finding the \gls{SDE} and second numerically solving it for calculating the desired particle locations, which typically leads to numerical issues due to various reasons, e.g., stiffness.
%
%
In contrast, the new method directly yields a closed-form \gls{ODE} by calculating the gradient and Hessian of the generalized Cram{\'e}r distance. This \gls{ODE} can be solved with standard methods.

\paragraph{Generality}
%
%
Methods relying on explicit mappings between the involved random variables are ultimately restricted to calculating posterior particle sets.
%
%
The reason is that calculating the flow for \glspl{GMD} or exponential densities based on the \gls{SDE} for the posterior random variable (associated to the Fokker-Planck equation) would add another layer of complication.
%
%
In fact, the \gls{SDE} is a dynamic system describing the evolution of the posterior random variable and thus, not even weight changes of a particle set are possible (as the system mapping only changes sample locations, but not sample weights).
%
%
In contrast, the new filtering method can be applied with different density representations.
See \cite{hanebeckProgressiveBayesNew2003} for an early approach involving \glspl{GMD}.

\section{Problem Formulation} \label{Sec_ProbForm}

%
%

\subsection{Models}

%
%
We consider a dynamic system with random state vector $\rvec{x}_k\in \IR^N$, defined at discrete time steps $k \in \IN \cup \{0\}$. The state evolves according to a nonlinear system equation
\begin{equation}
  \rvec{x}_{k+1} = \vec{a}_k(\rvec{x}_k, \vec{u}_k, \rvec{w}_k)
  \label{Eq_System_Equation}
\end{equation}
with nonlinear (deterministic) function $\vec{a}_k(\cdot, \cdot, \cdot)$, known input $\vec{u}_k$, and system noise $\rvec{w}_k$.

%
%
The state is usually hidden, i.e., $\rvec{x}_k$ is a latent variable not directly available, and is indirectly observed via measurements related to the state via the measurement equation
\begin{equation}
  \rvec{y}_k = \vec{h}_k(\rvec{x}_k) + \rvec{v}_k
  \label{Eq_Measurement_Equation}
\end{equation}
with nonlinear (deterministic) function $\vec{h}_k(\cdot)$ and measurement noise $\vec{v}_k$.

%
%
The models \Eq{Eq_System_Equation} and \Eq{Eq_Measurement_Equation} relate random variables and their realizations, and we call them generative models. Corresponding probabilistic models can be derived that describe the system evolution and the measurement extraction via conditional densities $f_k(\vec{x}_{k+1} | \vec{x}_k)$ and $f_k(\vec{y}_k | \vec{x}_k)$. For a specific measurement $\hat{\vec{y}}_k$, the so-called likelihood function $f_k^L$ is obtained by
\begin{equation}
  f_k^L(\vec{x})=f_k(\hat{\vec{y}}_k | \vec{x}_k)
  \enspace.
  \label{Eq_Likelihood}
\end{equation}

\subsection{Representation of Uncertain Latent States}

%
%
In the case of stochastic model and sensor uncertainties, the latent state is represented by a probability density function (pdf).
%
%
Continuous pdfs such as Gaussian or Gaussian mixture densities can be used in the case of linear systems. For nonlinear systems, however, the required transformations are difficult to handle for continuous densities, unless some form of linearization is employed, which limits the achievable quality.
%
%
Hence, often sample/particle representations are used.
%
%
In this paper, we employ deterministic sample representations we call Dirac mixtures, where the weights and locations are obtained by optimization rather than by random sampling from a continuous density.

\subsection{Estimation and Challenges}

%
%
We assume that the measurements $\hat{\vec{y}}_1$, $\hat{\vec{y}}_2$, $\cdots$ arrive sequentially and that we desire to provide a state estimate with each new measurement.
This requires the alternating execution of a prediction and a filter step.

%
%
The prediction step, or time update, entails propagating the estimated state density $f_k^e(\vec{x}_k)$ at time step $k$ to the next time step $k+1$ by using the generative description in equation \Eq{Eq_System_Equation} or the probabilistic description $f_k(\vec{x}_{k+1} | \vec{x}_k)$.
This results in the predicted state density $f_{k+1}^p(\vec{x}_{k+1})$.

%
%
The filter step, or measurement update, uses a measurement taken at time $k$ to correct the given prediction $f_k^p(\vec{x}_k)$ either based on the generative measurement equation in \Eq{Eq_Measurement_Equation} or the likelihood function \Eq{Eq_Likelihood}.

%
%
\begin{remark}
  Please note that in this paper, we focus on a single filter step, omit the time index $k$, and change $f_k^p(\cdot)$, $f_k^e(\cdot)$, $f_k^L(\cdot)$ to $f_p(\cdot)$, $f_e(\cdot)$, $f_L(\cdot)$.
\end{remark}

\subsection{Filter Step for Continuous Densities}

%
%
For a continuous prior density $f_p(\vec{x})$ and a continuous likelihood function $f_L(\vec{x})$,
the exact Bayesian filter step for calculating the desired continuous posterior density $f_e(\vec{x})$ is given by
\begin{equation}
  f_e(\vec{x}) = \frac{1}{c_e} \cdot f_p(\vec{x}) \cdot f_L(\vec{x})
  \label{Eq_FilterStepContinuous}
\end{equation}
with normalization constant
\begin{equation}
  c_e = \int_{\NewR^N} f_p(\vec{x}) \cdot f_L(\vec{x}) \, \d \vec{x}
  \label{Eq_NormConst}
\end{equation}
so that $\textstyle\int_{\NewR^N} f_e(\vec{x}) \, \d \vec{x} = 1$ holds.
%
%
Calculation of the normalization constant is in general a major computational burden.

\subsection{Challenges for General Density Representations}

%
%
The filter step in \Eq{Eq_FilterStepContinuous} is exact and assumes that the posterior density can completely capture the shape of the true distribution.
%
%
This is true, e.g., in linear systems.
%
%
For nonlinear systems, it only works when the posterior $f_e(\vec{x})$ lives in
a general infinite-dimensional distribution function space.
%
%
For densities with limited representation capabilities, i.e., finite mixtures or finite particle sets, as required in practical implementations, the density type changes and/or the complexity goes up.
%
%
For example, a \gls{GMD} becomes a non-Gaussian mixture during the update or an equally weighted particle set becomes an unequally weighted particle set.

%
%
In recursive Bayesian estimation, we desire the densities describing the considered latent variables to be closed under the measurement update step, at least in an approximate sense. This means that for a prior density of a certain type, we want the posterior density to be of the same type\footnote{Prior and posterior are then called (approximately) conjugate distributions w.r.t.\ a likelihood.}. In that case, we can use the resulting posterior as a prior for the next recursion step without changing its mechanism.
%
%
This means that we have to find the best approximation of the resulting posterior lying within the space of permissible densities.
This is equivalent to projecting the true posterior to the closest permissible approximation,
%
%
which requires a distance measure.

%
%
The optimization problem of finding the best approximative posterior density among the permissible densities is typically a complicated numerical minimization task that involves local minima. It is exacerbated by the fact that the prior is not a good initial guess, but usually the only available one.

\subsection{Filter Step for \glspl{DMD}}

%
%
In this paper, we focus on the case that the prior density $f_p(\vec{x})$ is not a continuous density, but given as a set of $L$ weighted samples (or particles).
It can be formally written as a \gls{DMD}
\begin{equation} \label{Eq_PriorDM}
  f_p(\vec{x}) = \sum_{i=1}^L w_{p,i} \cdot \delta(\vec{x}-\vec{x}_{p,i}) \enspace ,
\end{equation}
with weights $w_{p,i}>0$, $\textstyle\sum_{i=1}^L w_{p,i} = 1$ and sample locations $\vec{x}_{p,i}$.

%
%
For prior \glspl{DMD}, the posterior is also a \gls{DMD}, and we can omit the normalization constant from the Bayesian filter step \Eq{Eq_FilterStepContinuous}
\begin{equation}
  \tilde{f}_e(\vec{x}) \propto f_L(\vec{x}) \cdot f_p(\vec{x}) \enspace ,
  \label{Eq_BayesProp}
\end{equation}
as normalizing $\tilde{f}_e(\vec{x})$ becomes trivial.
%
%
The tilde $\tilde{\phantom{\ast}}$ marks this as the true posterior that will be approximated later.
%
%
Plugging in $f_p(\vec{x})$ from \Eq{Eq_PriorDM} gives
\begin{equation}
  \begin{aligned}
    \tilde{f}_e(\vec{x}) & \propto f_L(\vec{x}) \cdot \sum_{i=1}^L w_{p,i} \cdot \delta(\vec{x}-\vec{x}_{p,i})                                            \\
                         & = \sum_{i=1}^L \underbrace{w_{p,i} \cdot f_L(\vec{x}_{p,i})}_{\bar{w}_{e,i}} \cdot \, \delta(\vec{x}-\vec{x}_{p,i}) \enspace .
  \end{aligned}
  \label{Eq_OneShotPosteriorDM}
\end{equation}
Normalization to obtain the posterior weights is simply performed by
\begin{equation}
  \tilde{w}_{e,i} = \frac{\bar{w}_{e,i}}{\sum_{i=1}^L \bar{w}_{e,i}} \enspace.
\end{equation}
The posterior \gls{DMD} is now given as
\begin{equation}
  \tilde{f}_e(\vec{x}) = \sum_{i=1}^L \tilde{w}_{e,i} \cdot \delta(\vec{x}-\tilde{\vec{x}}_{e,i}) \enspace .
  \label{Eq_ExactPosteriorDM}
\end{equation}
%

\subsection{Challenges for \glspl{DMD}}

%
%
The posterior $\tilde{f}_e(\vec{x})$ in \Eq{Eq_ExactPosteriorDM} is derived from the straightforward application of the Bayesian filter step to Dirac mixtures.
%
%
Please note that in this case, only the weights change.
The posterior sample locations do not change w.r.t.\ the prior samples, i.e., $\tilde{\vec{x}}_{e,i} = \vec{x}_{p,i}$ for $i=1,\ldots,L$.
%
%
This leads to a serious problem:
The samples are not equally weighted anymore and do not equally contribute to the representation of the posterior.
Often, some particle weights are (close to) zero, in fact dying out, leading to particle degeneracy.
%
%
A typical scenario is large system noise, which spreads the particles during the prediction step combined with low measurement noise leading to ``narrow'' likelihoods.

%
%
Many solutions, some systematic, many of heuristic nature, have been proposed to solve the degeneracy problem,
%
%
which is a fundamental and difficult problem.

%
%
Our goal is to derive a Bayesian filter that inherently avoids degeneracy without any heuristic approaches.
It should be easy to understand, simple to implement, numerically stable, and robust.

\section{Homotopy Continuation} \label{Sec_Homotopy}

%
%
In order to circumvent the problems coming with the one-shot measurement update shown in the previous section, \cite{SPIE03_HanebeckBriechle-ProgBayes} proposed to gradually introduce the information from the measurement.
%
%
According to \cite{MFI03_Hanebeck}, this is done by carrying out the measurement update progressively over an artificial time $\gamma$, where $\gamma$ ranges in the (arbitrarily selected) interval%
\footnote{Other intervals could be used when more convenient. This also includes $\gamma \in [0, \infty]$.} $\gamma \in [0,1]$.

%
%
To achieve the desired effect, we parametrize the likelihood with the artificial time $\gamma$, resulting in $f_L(\vec{x},\gamma)$.
For the initial value of $\gamma$, $\gamma=0$, an uninformative likelihood, e.g., $f_L(\vec{x},\gamma=0)=1$, is selected that simply returns the prior unchanged.
For the final value of $\gamma$, $\gamma=1$, we want to recover the original likelihood function.
In between the extreme values for $\gamma$, i.e., $\gamma \in (0,1)$, we desire an interpolation between the constant likelihood and the original likelihood that is convenient to use and leads to the desired estimation results.

In summary, see also \cite[p.~8]{arXiv18_Hanebeck}, the progressive likelihood function $f_L(\vec{x},\gamma)$ parametrized by the artificial time $\gamma$ is given by
\begin{equation}
    f_L(\vec{x},\gamma) =
    \begin{cases}
        1                               & \gamma=0         \\
        \text{convenient interpolation} & \gamma \in (0,1) \\
        f_L(\vec{x})                    & \gamma=1
    \end{cases}
    \enspace .
\end{equation}

%
%
Plugging the parametrized likelihood into the Bayesian update equation \Eq{Eq_BayesProp} yields the true parametrized posterior
\begin{equation}
    \tilde{f}_e(\vec{x},\gamma) \propto f_L(\vec{x},\gamma) \, f_p(\vec{x}) \enspace .
    \label{Eq_ParametrizedPosterior}
\end{equation}

%
%
There are infinitely many progression schedules and, strictly speaking, a good schedule depends on the specific filtering setup, i.e., the prior density and the original likelihood.
%
%
An intuitive and generic progression schedule is attained by taking the power of the original likelihood as
\begin{equation}
    f_L(\vec{x},\gamma) = \left[ f_L(\vec{x}) \right]^{\gamma} \enspace .
\end{equation}
This obviously reduces to $f_L(\vec{x},\gamma=0) = 1$ and $f_L(\vec{x},\gamma=1) = f_L(\vec{x})$.
%
%
As typically most of the change in the posterior caused by changes in $\gamma$ is experienced for small $\gamma$, a generalization is to use nonlinear functions of $\gamma$ with $g(\gamma=0)=0$ and $g(\gamma=1)=1$ that start out with a smaller slope around zero such as $g(\gamma)=\gamma^2$.
The exponentiation with such a function gives a parametrized likelihood of the form
\begin{equation}
    f_L(\vec{x},\gamma) = \left[ f_L(\vec{x}) \right]^{g(\gamma)} \enspace .
\end{equation}
Its derivative is
\begin{equation}
    \begin{aligned}
        \dot{f}_L(\vec{x},\gamma)
         & = \frac{\partial}{\partial \, \gamma} \left\{ \left[
        \exp\Bigl( \log\bigl( f_L(\vec{x}) \bigr) \Bigr) \right]^{g(\gamma)} \right\}  \\[2mm]
         & = \frac{\partial}{\partial \, \gamma} \left\{
        \exp\Bigl( g(\gamma) \,\log\bigl( f_L(\vec{x}) \bigr) \Bigr) \right\}          \\[2mm]
         & = \log\bigl( f_L(\vec{x}) \bigr) \, \dot{g}(\gamma) \, f_L(\vec{x}, \gamma)
        \enspace .
    \end{aligned}
\end{equation}

%
%
In many applications, we assume additive Gaussian measurement noise, which leads to likelihoods from the exponential family.
%
%
For these likelihoods, we obtain simpler expressions due to the logarithm taken above. For additive zero-mean Gaussian noise with covariance matrix $\mat{C}_v$ and $g(\gamma)=\gamma$, the likelihood function corresponding to \Eq{Eq_Measurement_Equation} is
\begin{equation}
    f_L(\vec{x}, \gamma)
    = \exp\!\left( - \, \gamma \, \frac{1}{2}
    \bigl( \vec{y}-\vec{h}(\vec{x}) \bigr)^\top \mat{C}_v^{-1} \bigl( \vec{y}-\vec{h}(\vec{x}) \bigr)
    \right)
    \enspace .
\end{equation}

\section{Newton Flow Induced by Distance Measure} \label{Sec_NewtonFlow}

%
%
In a progressive Bayesian particle flow, a given intermediate posterior density undergoes continual changes for increasing artificial time $\gamma$ under the influence of the homotopic changes of the likelihood function.
%
%
Hence, we require a distance measure that matches a slightly (infinitesimally) changed posterior density with an approximation from the desired density class.
%
%
For particle sets, infinitesimal weight changes have to be instantly compensated by suitable particle location changes as we want to keep all weights equal.

%
%
We start with the definition of an abstract distance measure.
The \gls{ODE} for describing the flow will be derived in two steps.
In the first step, we will derive a so-called iterative flow where the given true posterior remains unchanged over the flow. Just the approximative posterior is changed by the flow.
In the second step, a so-called recursive \gls{ODE} is derived, where the reference density is replaced by the approximation at either discrete time steps or continually.

\subsection{Abstract Distance Measure}

%
%
We are given a true or reference posterior density $\tilde{f}_e(\vec{x},\gamma)$ explicitly depending on the artificial time $\gamma$
%
%
and an approximation $f_e(\vec{x}, \etavec(\gamma))$ of the reference density.
The parameter vector $\etavec(\gamma)$ is adjusted over time $\gamma$ so that the approximate posterior follows the evolving reference density.
As a result, $\etavec(\gamma)$ implicitly depends on $\gamma$.

%
%
To calculate the necessary changes in the parameter vector $\etavec(\gamma)$, an abstract distance measure for comparing $\tilde{f}_e(\vec{x},\gamma)$ and  $f_e(\vec{x}, \etavec(\gamma))$ is used
\begin{equation}
    D\bigl( \etavec(\gamma),\gamma \bigr)
    = D  \left(
    \tilde{f}_e(\vec{x},\gamma) ,
    f_e \left( \vec{x}, \etavec(\gamma) \right)
    \right)
    \enspace .
    \label{Eq_Distance}
\end{equation}
Changing $\gamma$ induces variations in the distance that have to be compensated by changes in $\etavec(\gamma)$ in order to stay in the minimum.

%
%
Finding the optimal approximate posterior density by minimizing the distance measure \Eq{Eq_Distance} is generally difficult due to local minima.
%
%
Hence, we start at the known minimum for $\gamma=0$, i.e., $f_e(\vec{x},\gamma=0)=\tilde{f}_e(\vec{x},\gamma=0)=f_p(\vec{x})$, and track the optimal $\gamma$-dependent approximate posterior $f_e(\vec{x}, \etavec(\gamma))$ as the minimum of \Eq{Eq_Distance} with infinitesimal changes of the artificial time $\gamma$.
%
%
The desired final approximate posterior is obtained at $\gamma=1$.

\subsection{Derivation of Iterative \gls{ODE}}

%
%
The necessary condition for a minimum is obtained by calculating the gradient vector $\vec{G}(\etavec(\gamma),\gamma)$, i.e., the partial derivative of $D(\etavec(\gamma),\gamma)$ w.r.t.\ parameter vector $\etavec(\gamma)$, and setting it to zero
\begin{equation}
    \vec{G}\bigl( \etavec(\gamma),\gamma \bigr)
    = \underbrace{\frac{\partial D(\etavec(\gamma),\gamma)}{\partial \etavec(\gamma)}}_{D_{\etavec}(\etavec(\gamma),\gamma)}
    = \vec{0} \enspace .
\end{equation}
As we are interested in the change of the parameter vector $\etavec(\gamma)$ w.r.t.\ $\gamma$, we take the partial derivative of $\vec{G}(\etavec(\gamma),\gamma)$ w.r.t.\ $\gamma$
\begin{equation}
    \underbrace{\frac{\partial \vec{G}(\etavec(\gamma),\gamma)}{\partial \etavec^T(\gamma)}}_{\vec{G}_{\etavec}\bigl( \etavec(\gamma),\gamma \bigr)}
    \cdot \frac{\partial \etavec(\gamma)}{\partial \gamma}
    + \underbrace{\frac{\partial \vec{G}\bigl( \etavec(\gamma),\gamma \bigr)}{\partial \gamma}}_{\vec{G}_{\gamma}(\etavec(\gamma),\gamma)}
    = \vec{0} \enspace
\end{equation}
which is equivalent to
\begin{equation}
    \underbrace{\frac{\partial^2 D\bigl( \etavec(\gamma),\gamma \bigr)}{\partial \etavec(\gamma) \, \partial \etavec^T(\gamma)}}_{D_{\etavec\etavec}(\etavec(\gamma),\gamma)}
    \cdot \underbrace{\frac{\partial \etavec(\gamma)}{\partial \gamma}}_{\dot{\etavec}(\gamma)}
    + \underbrace{\frac{\partial^2 D\bigl( \etavec(\gamma),\gamma \bigr)}{\partial \etavec(\gamma) \, \partial \gamma}}_{D_{\etavec\gamma}(\etavec(\gamma),\gamma)}
    = \vec{0} \enspace .
    \label{Eq_generalODE_with_braces}
\end{equation}
Using the abbreviations under the braces, we denote the derivative of the gradient vector w.r.t.\ $\gamma$ as
\begin{equation}
    \vec{J}\bigl( \etavec(\gamma), \gamma \bigr)
    = D_{\etavec\gamma}\bigl( \etavec(\gamma),\gamma \bigr)
    = \vec{G}_{\gamma}\bigl( \etavec(\gamma),\gamma \bigr)
\end{equation}
and the Hesse matrix as
\begin{equation}
    \mat{H}\bigl( \etavec(\gamma), \gamma \bigr)
    = D_{\etavec\etavec}(\etavec(\gamma),\gamma)
    =  \vec{G}_{\etavec}(\etavec(\gamma),\gamma)
    \enspace .
\end{equation}
We can now rewrite \Eq{Eq_generalODE_with_braces} and obtain the \gls{ODE}
\begin{equation}
    \mat{H}\bigl( \etavec(\gamma), \gamma \bigr)
    \cdot \dot{\etavec}(\gamma)
    + \vec{J}\bigl( \etavec(\gamma), \gamma \bigr)
    = \vec{0} \enspace .
    \label{Eq_ImplicitODE}
\end{equation}
%
%
It is an implicit \gls{ODE} with mass matrix $\mat{H}(\cdot,\cdot)$, where we have to solve for $\dot{\etavec}(\gamma)$.
%
%
It can be formally rewritten as an explicit \gls{ODE}
\begin{equation}
    \dot{\etavec}(\gamma)
    = - \mat{H}^{-1}\bigl( \etavec(\gamma), \gamma \bigr)
    \cdot \vec{J}\bigl( \etavec(\gamma), \gamma \bigr)
    \overset{\mathrm{def}}{=\joinrel=} \vec{R}\bigl( \etavec(\gamma), \gamma \bigr)
    \enspace .
    \label{Eq_ExplicitODE}
\end{equation}
%
%
Please note that we desist from explicitly inverting the Hessian to convert the \gls{ODE} into an explicit one, as numerical inversion has a high computational complexity and undesirable numerical properties. We rather solve for $\dot{\etavec}(\gamma)$ in the linear system \Eq{Eq_ImplicitODE} while exploiting the properties of the Hessian.

%
%
We call this a Newton flow for the parameter vector $\etavec(\gamma)$ as a continuous version of Newton's method, in analogy to a gradient flow being a continuous version of a gradient-descent method.

%
%
We also call this an iterative flow, as the reference posterior $\tilde{f}_e(\vec{x},\gamma)$ at time $\gamma$ is maintained as the product of the prior particle set times the progressive likelihood $f_L(\vec{x},\gamma)$.
%
%
As the product of prior density times likelihood typically is not in the same density class as the prior, the reference posterior $\tilde{f}_e(\vec{x},\gamma)$ strays into the territory of impermissible densities and, for significant changes in $\gamma$, creeps beyond the border of usefulness.
%
%
To make the effect on the flow in \Eq{Eq_ExplicitODE} more visible, we note that changes in $\gamma$ enter the right-hand side $\vec{R}(\cdot, \cdot)$ via the reference posterior $\tilde{f}_e(\vec{x},\gamma)$ and its derivative $\dot{\tilde{f}}_e(\vec{x},\gamma)$ w.r.t.\ $\gamma$. This means that we can write $\vec{R}(\cdot, \cdot)$ as an explicit function of $\tilde{f}_e(\vec{x},\gamma)$ and $\dot{\tilde{f}}_e(\vec{x},\gamma)$ as
\begin{equation}
    \dot{\etavec}(\gamma)
    = \vec{R}\bigl( \etavec(\gamma), \tilde{f}_e(\vec{x},\gamma), \dot{\tilde{f}}_e(\vec{x},\gamma) \bigr)
    \enspace .
\end{equation}
%

%
%
In order to avoid the reference posterior becoming useless, we have to replace it with its approximation $f_e(\vec{x}, \etavec(\gamma))$ either from time to time or continually.
We will call the resulting \gls{ODE} a recursive flow.

\subsection{Derivation of Recursive \gls{ODE}} \label{SubSec_generalRecFlow}

%
%
For deriving the recursive flow, we start with finite integration intervals%
\footnote{This is different from directly taking discrete steps in $\gamma$ as in \cite{Fusion16_Hanebeck} as we integrate the \gls{ODE} over the interval.}.
%
%
Imagine that we first integrate the \gls{ODE} from $\gamma=0$ to some small $\gamma_\ast$ so that the posterior flows from the prior to the best permissible approximation $f(\vec{x}, \etavec(\gamma_\ast))$ of the (impermissible) true posterior $\tilde{f}_e(\vec{x}, \gamma_\ast) \propto f_p(\vec{x}) f_L(\vec{x}, \gamma_\ast)$.
%
%
For \gls{DMD}, the weights of the reference posterior have changed and are now not equal anymore. To keep it from degenerating further, we replace the original reference posterior $\tilde{f}_e(\gamma_\ast)$ by its best approximation $f(\vec{x}, \etavec(\gamma_\ast))$ at time $\gamma_\ast$.
%
%
From this time on, we compare with this newly reset reference density, that we then simply call reference posterior for the next steps.
%
%
Of course, we have to modify the likelihood function that we use from $\gamma_\ast$ on as the new reference density already includes the likelihood integrated up to that time point.
%
%
This involves calculating an effective likelihood $f_L^\text{eff}(\vec{x},\gamma)$ by dividing the original likelihood $f_L(\vec{x}, \gamma)$ by the likelihood $f_L(\vec{x}, \gamma_\ast)$ used to far
\begin{equation}
    f_L^\text{eff}(\vec{x},\gamma) = \frac{f_L(\vec{x}, \gamma)}{f_L(\vec{x}, \gamma_\ast)}
    \enspace .
\end{equation}

%
%
We could now repeat this procedure for the next $\gamma$ interval (and in fact, the filtering method could be implemented this way).
%
%
However, we want to avoid the reference density leaving the space of permissible densities at all.
%
%
This is achieved by simply using infinitesimal integration intervals. This is equivalent to replacing the reference posterior by its best approximation for all $\gamma$ in the \gls{ODE}. Now the \gls{ODE} does not include the reference density at all anymore
\begin{equation}
    \dot{\etavec}(\gamma)
    = \vec{R}\bigl(
    \etavec(\gamma), f_L^\text{eff}(\vec{x},\gamma), \dot{f}_L^\text{eff}(\vec{x},\gamma)
    \bigr)
    \enspace .
\end{equation}
%
%
We call this a recursive flow.

\section{Generalized Cram{\'e}r Distance} \label{Sec_DistanceMeasure}

%
%
Instead of the abstract distance measure assumed in the previous section, we now use the specific distance measure in \Eq{Eq_CvD} from the Appendix that compares two \glspl{DMD} that may have (i)~different numbers of components $M$ and $L$, (ii)~at mutually distinct locations, (iii)~with non-equal weights.
%
%
For the particle flow in this paper, we use a specific configuration of this distance measure that compares two particle sets having the same numbers of samples ($M=L$).

\subsection{Adaptation of General Distance Measure to Particle Flow}

%
%
Let us define the given posterior density at artificial time $\gamma \in [0,1]$ as $\tilde{f}_e(\vec{x},\gamma)$.
%
%
When $\tilde{f}_e(\vec{x},\gamma)$ represents a particle set, the dependency on $\gamma$ stems from its $\gamma$-dependent weights.
%
%
This dependency is caused by the parametrized likelihood $f_L(\vec{x}, \gamma)$ as defined in \Sec{Sec_Homotopy} used for weighting the samples in the Bayesian update step.
From \Eq{Eq_OneShotPosteriorDM}, by plugging in $f_L(\vec{x}, \gamma)$, we obtain
\begin{equation}
    \tilde{f}_e(\vec{x},\tilde{\vec{w}}(\gamma), \tilde{\vec{X}})
    = \sum_{i=1}^L \tilde{w}_i(\gamma) \cdot \, \delta(\vec{x}-\tilde{\vec{x}}_i)
\end{equation}
with
\begin{equation}
    \tilde{\vec{X}} =
    \begin{bmatrix}
        \tilde{\vec{x}}_1^T, \tilde{\vec{x}}_2^T, \ldots, \tilde{\vec{x}}_L^T
    \end{bmatrix}^T
\end{equation}
and
\begin{equation}
    \tilde{\vec{w}}(\gamma) =
    \begin{bmatrix}
        \tilde{w}_1(\gamma), \tilde{w}_2(\gamma), \ldots, \tilde{w}_L(\gamma)
    \end{bmatrix}^T
    \enspace .
\end{equation}
We have constant $\tilde{\vec{x}}_i = \vec{x}_{p,i}$ and normalized time-varying weights
\begin{equation}
    \tilde{w}_i(\gamma)
    = \frac{f_L(\tilde{\vec{x}}_i, \gamma)}{F_L(\tilde{\vec{X}}, \gamma)}
    \label{Eq_normalizedWeightfromLikelihood}
\end{equation}
with
\begin{equation}
    F_L(\tilde{\vec{X}}, \gamma) = \sum_{i=1}^L f_L(\tilde{\vec{x}}_i,\gamma) \enspace .
    \label{Eq_totalLikelihood}
\end{equation}
%
%
It is obvious, that for $\gamma>0$, the weights $\tilde{w}_i(\gamma)$ become non-equal.

As we desire to maintain equal weights, we define an approximating posterior density
\begin{equation}
    f_e(\vec{x}, \vec{w}, \etavec(\gamma)) = \sum_{i=1}^L w_i \cdot \, \delta(\vec{x}-\vec{x}_i(\gamma))
\end{equation}
with fixed and equal weights $w_i = 1/L$ and $\gamma$-dependent locations $\vec{x}_i(\gamma)$. We collect the locations in a parameter vector $\etavec(\gamma)$ with
\begin{equation}
    \etavec(\gamma) =
    \begin{bmatrix}
        \vec{x}_1^T(\gamma), \vec{x}_2^T(\gamma), \ldots, \vec{x}_L^T(\gamma)
    \end{bmatrix}^T
\end{equation}
and the $\gamma$-independent weights in a vector
\begin{equation}
    \vec{w} =
    \begin{bmatrix}
        w_1, w_2, \ldots, w_L
    \end{bmatrix}^T
    \enspace .
\end{equation}
%

%
%
We define the distance $D$ between the true density $\tilde{f}_e$ and its approximation $f_e$ based on the distance measure \Eq{Eq_CvD} as
\begin{equation}
    D(\etavec(\gamma),\gamma)
    = D \Bigl(
    \tilde{f}_e(\vec{x},\tilde{\vec{w}}(\gamma), \tilde{\vec{X}}) ,
    f_e \bigl( \vec{x}, \vec{w}, \etavec(\gamma) \bigr)
    \Bigr)
    \enspace .
\end{equation}

%
%
The best approximation of the posterior that maintains equal weights is tracked as the minimum of $D$ over artificial time $\gamma$, starting with the posterior density for $\gamma=0$, i.e., $f_e \bigl( \vec{x}, \vec{w}, \etavec(0) \bigr)=f_p(\vec{x})$.

%
%
\subsection{\gls{ODE} for Iterative Particle Flow}

%
%
During the particle flow, the densities $\tilde{f}_e$ and $f_e$ have the same number of components, so we have $M=L$.
In that case, the gradient in \Eq{Eq_Gradient} from the Appendix simplifies to
\begin{equation}
    \begin{aligned}
        \vec{G}_k = & \frac{\partial D}{\partial \vec{x}_k}                                               \\
        =           & 4 \, w_k \Bigg\{
        \sum_{\substack{i=1                                                                               \\ i \neq k}}^L w_i \, \vec{\mathcal{G}}(\vec{x}_k - \vec{x}_i)
        - \sum_{i=1}^L \tilde{w}_i \, \vec{\mathcal{G}}(\vec{x}_k - \tilde{\vec{x}}_i)
        \Bigg\}                                                                                           \\
                    & + 2 \, K \, w_k \sum_{i=1}^L (w_i \, \vec{x}_i  - \tilde{w}_i \, \tilde{\vec{x}}_i)
    \end{aligned}
\end{equation}

%
%
We note that $\tilde{w}_i$ is a function of $\gamma$, i.e., $\tilde{w}_i = \tilde{w}_i(\gamma)$, while $\tilde{\vec{x}}_i$ and $w_i$ are independent of $\gamma$, and $\etavec_i$ only implicitly depends on $\gamma$.
%
%
The derivative of the gradient vector w.r.t.\ $\gamma$ is given by
\begin{equation}
    \vec{J}(\gamma) = \frac{\partial \vec{G}(\gamma)}{\partial \gamma}
    = \begin{bmatrix}
        \vec{J}_1^T(\gamma), \vec{J}_1^T(\gamma), \ldots, \vec{J}_L^T(\gamma)
    \end{bmatrix}^T
    \label{Eq_dG_dgammaPF}
\end{equation}
with elements
\begin{equation}
    \begin{aligned}
        \vec{J}_k(\gamma) = & \frac{\partial \vec{G}_k(\gamma)}{\partial \gamma} \\
        =                   & - w_k \sum_{i=1}^L
        \bigg\{
        4 \, \dot{\tilde{w}}_i(\gamma) \, \vec{\mathcal{G}}(\vec{x}_k - \tilde{\vec{x}}_i)
        + 2 \, K \dot{\tilde{w}}_i(\gamma) \, \tilde{\vec{x}}_i
        \bigg\} \enspace .
    \end{aligned}
\end{equation}

%
%
The normalized weights $\tilde{w}_i(\gamma)$ induced by the parametrized likelihood $f_L(\vec{x}, \gamma)$ are given in
\Eq{Eq_normalizedWeightfromLikelihood} and \Eq{Eq_totalLikelihood}.
%
%
The derivatives of the normalized weights w.r.t.\ artificial time $\gamma$ are
\begin{equation}
    \dot{\tilde{w}}_i(\gamma) = \frac{\dot{f}_L(\tilde{\vec{x}}_i, \gamma) \, F_L(\tilde{\vec{X}}, \gamma)
        - f_L(\tilde{\vec{x}}_i, \gamma) \, \dot{F}_L(\tilde{\vec{X}}, \gamma)}{F_L^2(\tilde{\vec{X}}, \gamma)}
\end{equation}
for $i=1,2,\ldots,L$ with
\begin{equation}
    \dot{F}_L(\tilde{\vec{X}}, \gamma) = \sum_{i=1}^L \dot{f}_L(\tilde{\vec{x}}_i, \gamma)
    \enspace .
\end{equation}
%

\subsection{\gls{ODE} for Recursive Particle Flow}

%
%
In analogy to the procedure in \SubSec{SubSec_generalRecFlow}, we now derive a recursive version of the particle flow.
%
%
This leads to simplifications of the derivative of the gradient vector w.r.t.\ $\gamma$ from \Eq{Eq_dG_dgammaPF}. Its elements are now given by
\begin{equation}
    \vec{J}_k(\gamma) = - w_k \sum_{i=1}^L
    \bigg\{
    4 \, \dot{w}_i(\gamma) \, \vec{\mathcal{G}}(\vec{x}_k - \vec{x}_i)
    + 2 \, K \dot{w}_i(\gamma) \, \vec{x}_i
    \bigg\} \enspace .
\end{equation}

%
%
The derivatives of the normalized weights w.r.t.\ artificial time $\gamma$ are simplified to
\begin{equation}
    \dot{w}_i(\gamma) = \frac{1}{L}
    \bigg\{
    \frac{\dot{f}_L(\vec{x}_i, \gamma)}{f_L(\vec{x}_i, \gamma)}
    - \frac{1}{L} \sum_{j=1}^L \frac{\dot{f}_L(\vec{x}_j, \gamma)}{f_L(\vec{x}_j, \gamma)}
    \bigg\}
\end{equation}
for $i=1,2,\ldots,L$.

%
%
The Hesse matrix in \Eq{Eq_Hessian} from the Appendix is now also simplified.
%
%
The diagonal block matrices are given by
\begin{equation}
    \mat{H}_{kk} = 2 \, w_k^2 \, \bigl\{ K - 2 \, \log\bigl(d_\text{min}^2\bigr) \bigr\} \, \mat{I}_N
\end{equation}
for $k = 1,\ldots, L$
%
%
and the off-diagonal block matrices are still given by
\begin{equation}
    \mat{H}_{kl} = 2 \, w_k \, w_l \, \big\{ K \, \mat{I}_N
    - 2 \, \bm{\mathcal{H}}(\vec{x}_k - \vec{x}_l) \big\}
\end{equation}
for $k = 1,\ldots, L$, $l=1,\ldots,L$, $l \neq k$.

\section{Conclusions} \label{Sec_Conclude}

%
%
The proposed \gls{NFPF} is a \emph{systematic approach} to recursively estimate the hidden state of a nonlinear dynamic system. No heuristics are involved.
%
%
The \gls{NFPF} is \emph{easy to understand and simple to use} as almost no tuning factors are required.
%
%
It is \emph{fast} for several reasons: (i)~The number of particles to be processed is much smaller than in a standard particle filter. (ii)~The distance measure, its gradient, and the Hesse matrix can be calculated in closed form. (iii)~No optimization is required, just the solution of an \gls{ODE} over a finite time interval.
%
%
The proposed estimator shows a \emph{high performance} compared to state-of-the-art approaches,
produces unbiased high-quality estimates, is robust, is degeneration-free by design, and works in high dimensions.
%
%
The proposed plugin replacement for a particle filter step is \emph{fully differentiable}.
This allows for end-to-end learning of the measurement nonlinearities or the noise densities, which will be exploited in future work.

\section*{Appendix} \label{Sec_Appendix}

%
%
For convenience, we summarize the set distance measure derived in \cite{AT15_Hanebeck,arXiv14_Hanebeck-Reduction} and list all the required formulas in the slightly different notation used in this paper.

\subsection{Distance Measure}
%
%
We are given two \glspl{DMD} that we want to compare.
Usually, this comparison is used for the approximation of one \gls{DMD} by the other.
%
%
The reference \gls{DMD} is called $\tilde{f}(\vec{x})$.
It comprises $M$ components with weights $\tilde{w}_1, \tilde{w}_2, \ldots, \tilde{w}_M$ and locations
\begin{equation}
    \tilde{\vec{x}}_j
    = \begin{bmatrix}
        \tilde{x}_{1,j}, \tilde{x}_{2,j}, \ldots, \tilde{x}_{N,j}
    \end{bmatrix}^T \in \NewR^N
\end{equation}
for $j=1,\ldots, M$.
%
%
The approximating \gls{DMD} $f(\vec{x})$ has $L$ components with weights $w_1, w_2, \ldots, w_L$ and locations
\begin{equation}
    \vec{x}_i = \begin{bmatrix} x_{1,i}, x_{2,i}, \ldots, x_{N,i} \end{bmatrix}^T \in \NewR^N
\end{equation}
for $i=1,\ldots, L$.

%
%
The numbers of components $L$ and $M$ can be different. In reduction problems, the number of components of the approximating density is usually less than or equal to the number of components $M$ of the reference density.
In some cases, we even have $L \ll M$.
%
%
For the application of the distance in this paper, the numbers of components are equal, i.e., $M=L$.

%
%
The generalized Cram{\'e}r-von Mises distance \cite{AT15_Hanebeck} is given by $\sqrt{D}$ with
\begin{equation}
    D = D\bigl( \tilde{f}, f \bigr)
    = D_{\tilde{x}\tilde{x}} - 2 D_{x\tilde{x}} + D_{xx} + K \, D_{E} \enspace ,
    \label{Eq_CvD}
\end{equation}
with
\begin{equation}
    D_{\tilde{x}\tilde{x}} = \sum_{i=1}^M \sum_{j=1}^M \tilde{w}_i \, \tilde{w}_j \,
    \mathcal{D}(\tilde{\vec{x}}_i - \tilde{\vec{x}}_j)
    \enspace ,
\end{equation}
\begin{equation}
    D_{x\tilde{x}} = \sum_{i=1}^L \sum_{j=1}^M w_i \, \tilde{w}_j \,
    \mathcal{D}(\vec{x}_i - \tilde{\vec{x}}_j)
    \enspace ,
\end{equation}
\begin{equation}
    D_{xx} = \sum_{i=1}^L \sum_{j=1}^L w_i \, w_j \,
    \mathcal{D}(\vec{x}_i - \vec{x}_j)
    \enspace ,
\end{equation}
and
\begin{equation}
    \mathcal{D}(\vec{d}) = \vec{d}^T \vec{d} \, \log\bigl(d_\text{min}^2 + \vec{d}^T \vec{d}\bigr)
    \enspace .
\end{equation}
$D_{E}$ can be viewed as a penalty term that ensures equal means and is given by the squared difference between the means of $\tilde{f}(\vec{x})$ and $f(\vec{x})$
\begin{equation}
    D_{E} = \left\| \sum_{i=1}^L w_i \, \vec{x}_i
    - \sum_{j=1}^M \tilde{w}_j \, \tilde{\vec{x}}_j \right\|_2^2
    \enspace .
\end{equation}
%
%
$K$ is a large constant for weighting the penalty term.

%
%
Please note that in contrast to \cite{AT15_Hanebeck}, we added the small constant $d_\text{min}$ here to avoid singularities when two samples assume the same location.

\subsection{Gradient}

The gradient of $D$ in \Eq{Eq_CvD} w.r.t.\ $\vec{x}_k$ is given by%
\footnote{For minimization, we use the squared distance $D$ for simplicity.}
\begin{equation}
    \begin{aligned}
        \vec{G}_k = & \frac{\partial D}{\partial \vec{x}_k} \\
        =           & 4 \, w_k \Bigg\{
        \sum_{\substack{i=1                                 \\ i \neq k}}^L w_i \, \vec{\mathcal{G}}(\vec{x}_k - \vec{x}_i)
        - \sum_{j=1}^M \tilde{w}_j \, \vec{\mathcal{G}}(\vec{x}_k - \tilde{\vec{x}}_j)
        \Bigg\}                                             \\
                    & + 2 \, K \, w_k \Bigg\{
        \sum_{i=1}^L w_i \, \vec{x}_i
        - \sum_{j=1}^M \tilde{w}_j \, \tilde{\vec{x}}_j
        \Bigg\}
    \end{aligned}
    \label{Eq_Gradient}
\end{equation}
with
\begin{equation}
    \vec{\mathcal{G}}(\vec{d}) = \left[
        \frac{\vec{d}^T \vec{d}}{d_\text{min}^2 + \vec{d}^T \vec{d}}
        + \log\bigl(d_\text{min}^2 + \vec{d}^T \vec{d}\bigr)
        \right] \, \vec{d}
    \enspace .
\end{equation}

\subsection{Hesse Matrix}

%
%
The Hesse matrix $\mat{H}$ of the distance measure $D$ is composed of blocks $\mat{H}_{kl} \in \NewR^{N \times N}$ for $k = 1,\ldots, L$, $l=1,\ldots,L$ according to
\begin{equation}
    \mat{H} =
    \begin{bmatrix}
        \mat{H}_{11} & \mat{H}_{12} & \mat{H}_{13} & \cdots & \mat{H}_{1L} \\
        \mat{H}_{21} & \mat{H}_{22} & \mat{H}_{23} & \cdots & \mat{H}_{2L} \\
        \vdots       & \vdots       & \vdots       & \ddots & \vdots       \\
        \mat{H}_{L1} & \mat{H}_{L2} & \mat{H}_{L3} & \cdots & \mat{H}_{LL}
    \end{bmatrix}
    \enspace .
    \label{Eq_Hessian}
\end{equation}
%
%
We have to distinguish between diagonal and off-diagonal block matrices.
%
%
The diagonal block matrices are given by
\begin{equation}
    \begin{aligned}
        \mat{H}_{kk} & = \frac{\partial G_k}{\partial \vec{x}_k^T}
        = \frac{\partial^2 D}{\partial \vec{x}_k \partial \vec{x}_k^T} \\
                     & = 2 \, K \, w_k^2 \, \mat{I}_N                  \\
                     & + 4 \, w_k \Bigg\{
        \sum_{\substack {i=1                                           \\ i \neq k}}^L
        w_i \, \bm{\mathcal{H}}(\vec{x}_k - \vec{x}_i)
        - \sum_{j=1}^M \tilde{w}_j \, \bm{\mathcal{H}}(\vec{x}_k - \tilde{\vec{x}}_j)
        \Bigg\}
    \end{aligned}
\end{equation}
for $k = 1,\ldots, L$ with
\begin{equation}
    \begin{aligned}
        \bm{\mathcal{H}}(\vec{d}) & =
        2 \frac{2 \, d_\text{min}^2 + \vec{d}^T \vec{d}}{(d_\text{min}^2 + \vec{d}^T \vec{d})^2}
        \, \vec{d} \, \vec{d}^T              \\
                                  & + \left[
            \frac{\vec{d}^T \vec{d}}{d_\text{min}^2 + \vec{d}^T \vec{d}}
            + \log\bigl(d_\text{min}^2 + \vec{d}^T \vec{d}\bigr)
            \right] \, \mat{I}_N
        \enspace ,
    \end{aligned}
\end{equation}
where $\mat{I}_N$ is the $N$-dimensional identity matrix.
%
%
The off-diagonal block matrices are given by
\begin{equation}
    \begin{aligned}
        \mat{H}_{kl} & = \frac{\partial G_k}{\partial \vec{x}_l^T}
        = \frac{\partial^2 D}{\partial \vec{x}_k \partial \vec{x}_l^T} \\
                     & = 2 \, w_k \, w_l \, \big\{ K \, \mat{I}_N
        - 2 \, \bm{\mathcal{H}}(\vec{x}_k - \vec{x}_l) \big\}
    \end{aligned}
\end{equation}
for $k = 1,\ldots, L$, $l=1,\ldots,L$, $l \neq k$.
%
%
The Hesse matrix is symmetric and positive definite.


\printbibliography

\end{document}